\documentclass[12pt]{article}

\usepackage{graphicx} 
\usepackage{{realboxes}}
\graphicspath{{subtex/img/}} 
\usepackage{caption}
\usepackage{subcaption}
\usepackage{comment}
\usepackage{geometry}
\usepackage{pdflscape}

\usepackage{algorithm}
\usepackage{algpseudocode}
\makeatletter
\let\OldStatex\Statex
\renewcommand{\Statex}[1][3]{%
	\setlength\@tempdima{\algorithmicindent}%
	\OldStatex\hskip\dimexpr#1\@tempdima\relax}
\makeatother

\usepackage{booktabs} 
\usepackage{threeparttable} 
\usepackage{tabularx}
\usepackage{amsmath,amsfonts,amsthm,bm} 
\usepackage{soul}
\usepackage{siunitx}
\usepackage{eurosym}
\usepackage{tikz}

\usepackage{bbm}

\usepackage{scalerel}

\usepackage{natbib}
\usepackage[blocks]{authblk}

\usepackage[hidelinks]{hyperref}

\usepackage[capitalise,noabbrev]{cleveref}

\newtheorem{proposition}{Proposition}

\Crefname{hypothesis}{Hypothesis}{Hypotheses}
\Crefname{lemma}{Lemma}{Lemmata}
\Crefname{proposition}{Proposition}{Propositions}
\Crefname{assumption}{Assumption}{Assumptions}

\begin{document}

\begin{titlepage}
    \newgeometry{top=3cm,bottom=3cm}
    
	\title{When Should we Expect Non-Decreasing Returns from Data in Prediction Tasks?
}
	
	\vspace{1cm}
	\author{
	\begin{minipage}{0.99\textwidth}
	\centering
	Maximilian Schaefer\footnote{Institut Mines-Télécom Business School, \href{mailto:maximilian.schafer@imt-bs.eu}{maximilian.schafer@imt-bs.eu}.
	The author is also affiliated with Universit\'e Paris-Saclay, Universit\'e d'\'Evry, and the LITEM Laboratory.} 
	\end{minipage}
	}
	\date{\today}
	\maketitle
	
\vspace{1cm}

\begin{abstract}

This article studies the change in the prediction accuracy of a response variable when the number of predictors increases, and all variables follow a multivariate normal distribution. Assuming that the correlations between variables are independently drawn, I show that adding variables leads to globally increasing returns to scale when the mean of the correlation distribution is zero. The speed of learning depends positively on the variance of the correlation distribution. I use simulations to study the more complex case of correlation distributions with a non-zero mean and find a pattern of decreasing returns followed by increasing returns to scale — as long as the variance of correlations is not degenerate, in which case globally decreasing returns emerge. I train a collaborative filtering algorithm using the MovieLens 1M dataset to analyze returns from adding variables in a more realistic setting and find globally increasing returns to scale across $2,000$ variables. The results suggest significant scale advantages from additional variables in prediction tasks.


\vspace{1cm}


\end{abstract}
\setcounter{page}{0}
\thispagestyle{empty}
\end{titlepage}

\section{Introduction}
\label{sec:intro}

Data, as an input into prediction tasks, may either come in the form of additional observations ($N$), additional variables ($K$), or both simultaneously.\footnote{For example, when a user rates an item in a recommendation system, she not only increases the observation count on that item by one, but she also simultaneously increases the vector of observable characteristics about herself.} While the statistical theory of large $N$ asymptotics provides a theoretical underpinning for the hypothesis that data have decreasing returns to scale \citep{bajari2019impact}, it only covers the $N$-dimension of data. By contrast, there seems to be no solid understanding about the nature of returns for the $K$-dimension of data, despite emerging empirical evidence pointing toward return properties that deviate from the $N$-dimension.          

To provide insights into the nature of returns in the $K$-dimension, I model learning from additional variables as the conditional variance reduction of a target variable $y$ when increasing its conditioning set $\mathbf{x_k}$. I assume that variables follow a multivariate standard normal distribution with a positive definite correlation matrix, where correlations are independently drawn from a probability distribution with finite second moments.\footnote{Assuming a standard normal distribution instead of a normal distribution is without loss of generality. To keep the exposition concise and avoid notational clutter, the main text only refers to the multivariate standard normal distribution. However, the proof of the main result in \cref{app:proof} covers the more general case of the multivariate normal distribution.} The correlations determine the conditional variance of $y$, which corresponds to the mean squared error (MSE) of an ordinary least squares (OLS) regression of $y$ on $\mathbf{x_k}$. My results characterize the behavior of OLS predictions that would emerge when $K$ increases and $N$ is very large.\footnote{The large $N$ assumption is implicit in the absence of an estimation error in the model and allows me to focus the research question on $K$.}

I show that the MSE is a strictly decreasing and concave function of the number of predictor variables when the correlation distribution has a mean of zero, which implies globally increasing returns to scale in prediction accuracy. Since the MSE is bounded below by zero, the previous statement naturally only applies to the case in which this bound has not yet been reached. The overall speed of learning is governed by the variance of the correlation distribution, with a higher variance leading to faster learning. To study the case of correlation distributions with a mean different from zero, which is significantly more complex to analyze theoretically, I use simulations and draw correlations from a truncated Student's t-distribution with a positive mean. 

Across different parameterizations of the Student's t-distribution, I find a typical pattern of first decreasing and then increasing returns to scale as variables accumulate. I discuss the mechanism behind this result, which is related to the average collinearity between predictor variables when the mean of the correlations is different from zero. This collinearity first leads to diminishing returns. However, as predictors continue to accumulate, the collinearity between predictors is projected out, and the same dynamics as for the mean-zero case eventually take over. The only exception is when the correlation distribution is nearly degenerate (variance near zero), in which case globally decreasing returns to scale materialize.

The simulations help illustrate the setting studied in this article. The fact that correlations between variables follow a certain distribution appears natural, as the information that can be learned from one variable about another should exhibit variation across variable pairs. Assuming independent draws from this distribution amounts to the assumption that the precise sequence in which a firm accumulates variables about users cannot be controlled. I consider this an approximation of the scenario where the variables observed about a user depend on her choices, and these choices are determined to a large degree by her idiosyncratic tastes. This is typical in recommendation systems where a user's idiosyncratic interaction with items determines the characteristics observed about her.

One simulated MSE trajectory can be thought of as the prediction quality a new user experiences if she joins a firm with a personalized recommendation system. The trained parameters of the recommendation system are modeled by the correlation coefficients between variables. The firm can estimate these correlations with an arbitrary degree of accuracy as it benefits from a large number of existing users and, hence, observations (i.e., large-$N$ asymptotics are fulfilled). As the new user starts interacting with the system, she generates variables that allow the recommendation engine to input these variables into its pre-trained algorithm. The average MSE trajectory corresponds to the average prediction accuracy the user will experience across the different possible sequences of variable accumulation. Alternatively, and perhaps more straightforwardly, the setting models the scenario where a firm with many users acquires new variables about every user and updates its recommendation model using these new variables. In this scenario, the individual $R$-squared trajectories can be thought of as the average improvement experienced across users for one possible combination of variables added.\footnote{The conditional variance does not depend on the specific realizations of the random variable in the conditioning set. For the conditional variance of $y$ given $x$, only the correlation between both matters, not the specific value $x$ takes. As a consequence, we can interpret the conditional variance as contingent on a particular $x$ realization or as an average across users' $x$ realizations.}

I also use simulations to assess the robustness of the findings to fat-tailed correlation distributions by drawing correlations from a Student's t-Distribution with a degrees of freedom parameter (DoF) of one. This scenario covers the case where most variables have a small expected contribution to improving the prediction, with some potentially very large outliers that can suddenly significantly improve the prediction accuracy. While the results suggest that fat tails reduce the concavity, they do not reverse the result, and hence, increasing returns continue to be observed in the same manner as with thin-tailed distributions. 

To assess the external validity of the predictions generated within the stylized framework of the multivariate normal, I train a collaborative filtering algorithm \citep{Hug2020} on the MovieLens 1M dataset \citep{harper2015movielens} and assess the RMSE of its prediction on a hold-out sample. The theoretical framework suggests analyzing how the RMSE changes when a same set of new movies is rated by \textit{all} users. In practice, this is impossible to implement as no single movie is rated by all users. Instead, I evaluate the RMSE as the number of movies with at least 100 ratings increases. I average my results across different movie accumulation scenarios and find increasing returns to scale. Notably, increasing returns are observed after $2,000$ variables have been added, which is the maximum that can be considered given the data. This suggests that the region of increasing returns is large, consistent with the model predictions based on the multivariate normal distribution.

This article contributes to the empirical literature on returns to data in economics and management. One strand treats the $N$-dimension of data and finds evidence consistent with decreasing returns to scale: \cite{chiou2017search} exploit changes in the retention policy of search engine logs to causally analyze the effect of historical data on result accuracy. \cite{bajari2019impact} analyze potential interaction effects of different data dimensions in the context of Amazon's sales forecast system.\footnote{Their setting differs from mine in important ways. In particular, it does not map onto the recommendation system setting and assumes a fixed set of observed and unobserved explanatory variables driving common patterns across different prediction tasks.} \cite{klein2022important} conduct an experiment to assess the effect of additional observations on the quality of search results, and \cite{peukert2024editor} study the effect of prior user visits on prediction accuracy in the context of news recommendation. 

An explicit analysis of the $K$-dimension is provided in the empirical section of \cite{lee2021recommender}, who find decreasing returns to scale in a recommendation system context. Analyzing both the $N$- and the $K$-dimension simultaneously, \cite{schaefer2023complementarities} find evidence consistent with increasing returns to scale in the $K$-dimension. A similar result is found in \cite{yoganarasimhan2020search}, who concludes that long-term personalization is more valuable than short-term personalization. \cite{ullrich2024returns} note that the decreasing returns in the $K$-dimension are less pronounced than the decreasing returns in the $N$-dimension. \cite{hocuk2022economies} is the only empirical article whose setting closely corresponds to the theoretical framework described here. The authors use health record data to train machine learning algorithms based on a varying number of variables, $K$, while holding the number of observations, $N$, fixed. They assess the performance of the machine learning algorithms in predicting health outcomes and find evidence consistent with increasing returns to scale in $K$. In independent work, \cite{colla2025theory} shows increasing returns to scale from additional variables in a Bayesian model with uncorrelated predictors and a maximum entropy prior. By contrast, I study a non-Bayesian setting with possibly correlated predictors which might generate decreasing returns, and provide evidence for the external validity of my results. 

The findings presented in this article may help rationalize some of the conflicting findings in the existing literature analyzing the $K$-dimension. Since first decreasing and then increasing returns to scale might occur, there is a risk that empirical studies might not include enough variables to reach the part of the learning curve that exhibits increasing returns. This could be the case in \cite{lee2021recommender}, who consider only nine conditioning variables. Additionally, the simulations reveal that decreasing returns across many variables, as documented by \cite{ullrich2024returns}, are possible when predictors are collinear \textit{and} if the correlation distribution has a low variance. Thus, besides providing a theoretical underpinning for the possibility of increasing returns in the $K$-dimension, this article contributes to the literature by generating insights into the precise determinants of the returns.

This article is also related to the literature on the value of information, particularly to \cite{radner1984nonconcavity}, who provide conditions for non-concavity in the value of information in the context of Bayesian decision problems, a subject further analyzed in \cite{chade2002another} and \cite{keppo2008demand}. \cite{azevedo2020b} study how to allocate scarce observations ($N$) across experiments and show that the non-concavity result hinges on thin-tailed priors and breaks down under fat tails. The aforementioned studies analyze optimal economic decision-making based on plausible, empirically founded assumptions about prediction accuracy.

By contrast, this article provides novel insights into how prediction accuracy changes when increasing the number of variables. It can be viewed as analyzing the properties of a data production function that takes variables as input and outputs prediction quality. Thus, this article contributes to understanding which model assumptions are realistic depending on the context studied. The findings reveal that the paradigm of decreasing returns to scale from data, which has a sound statistical underpinning in the $N$-dimension, might not hold for the $K$-dimension. As such, my findings help inform modeling decisions in economic analyses of data-driven markets and the competition policy debate surrounding the value of data.\footnote{A non-exhaustive list of the economic and management literature analyzing market outcomes in data-driven markets includes \cite{farboodi2019big, pruefer_schottmueler_comp_big_data_2017, ref:hagiu2020, carballa2025data}, and \cite{calzolari2025machine}. Also related is the literature on data externalities \citep{acemoglu2019too, bergemann2021economics, aguiar2022facebook}, for which my results suggest super-linear increases. With respect to the literature surrounding the “big data” antitrust debate, relevant references include \cite{lerner2014role, newman_search_2014, shepp_wambach_bigdata_mp_2015, stucke_grunes_bigdata_anitrust_2015, lambrecht2017can_nv, sokol_comerford_bigdata_antitrust_2017}, and \cite{tucker_platforms_comp_2019}.}

My findings are also relevant for the question of how improvements in prediction technology map into economic productivity. On a microeconomic level, several studies have found productivity gains from AI across different tasks \citep{noy2023experimental, cui2024effects, hoffmann2024generative}. On a macroeconomic level, \cite{acemoglu2025simple} suggests modest effects of AI on total factor productivity. By contrast, \cite{merali2024scaling} provides experimental evidence for scaling between AI performance and labor productivity, suggesting effects an order of magnitude larger, underscoring the importance of accounting for scaling properties in economic models. 

A large body of literature from computer science investigates the scaling laws between data and prediction loss. Typically, this literature finds decreasing returns to scale in the number of training samples, i.e., $N$ \citep{kaplan2020scaling}.\footnote{I am not aware of a study from computer science that conceptualizes learning from data in a similar way to this article.} It has been noted that the empirically observed returns decrease significantly less fast than predicted by statistical theory \citep{hestness2017deep}. The findings of this article suggest the possibility that this might be a consequence of not accounting for the $K$-dimension, which is not always trivial to identify. For example, distinguishing a variable from an observation is clear in the context of OLS, but much less so in, for example, image recognition with neural networks. On a more abstract level, $K$ might be seen as the number of relevant features for a prediction problem, which is not always easily measurable, especially with sophisticated algorithms that can be thought of as learning features autonomously. It is beyond the scope of this article to explore this question in detail. The collaborative filtering algorithm used in the empirical section allows operationalizing the distinction between variables and observations. Nevertheless, the findings presented here suggest that the distinction between variables (features) and observations (samples) might merit more attention to better understand data returns, also with sophisticated algorithms.

In the remainder of the article, \cref{sec:model} briefly introduces notation and recapitulates the theoretical framework used to study returns from adding variables. \cref{sec:results_I} presents the results for a multivariate normal distribution: \cref{subsec:mean_zero} summarizes the theoretical result obtained when the correlation distributions have a mean of zero. \cref{subsec:mean_nzero} presents the simulation results used to analyze the case of correlation distributions with a mean different from zero and provides a heuristic argument for why the observed patterns should hold generally. \cref{subsec:discussion} provides a more in-depth discussion of the model assumptions in light of the obtained results. \cref{sec:results_II} covers the empirical analysis based on the MovieLens 1M Dataset: \cref{subsec:data} introduces the data, \cref{subsec:svd} the collaborative filtering algorithm, \cref{subsec:sim} details the empirical strategy, and \cref{subsec:simres} presents the results. Finally, \cref{sec:conc} concludes the article.

\section{The Multivariate Normal Distribution}
\label{sec:model}

I study learning from variables by assuming that the $K$-dimensional random vector $\mathbf{x} = (x_1, \ldots, x_K)$ follows a multivariate standard normal distribution with mean vector $\bm{\mu} = 0$ and a random realization of a positive definite correlation matrix $\bm{\Sigma}$.\footnote{Assuming a standard normal distribution instead of a normal distribution is without loss of generality. To keep the exposition concise and avoid notational clutter, the main text only refers to the multivariate standard normal distribution. However, the proof of the main result in \cref{app:proof} covers the more general case of the multivariate normal distribution.} Using an arbitrary subset of variables $\mathbf{x_k} \in \mathbf{x}$ as the predictors for a target variable, $y$, selected from the complementary set $\mathbf{x} \backslash \mathbf{x_k}$, the conditional variance of $y$ given $\mathbf{x_k}$ is: 

\begin{equation}\label{eq:multnorm}
	\sigma^2_{y|\mathbf{x_k}} = 1 - \Sigma_{y,\mathbf{x_k}}(\Sigma_{\mathbf{x_k},\mathbf{x_k}})^{-1}\Sigma_{y,\mathbf{x_k}} = 1 - R^2_{y|\mathbf{x_k}}.
\end{equation}

\noindent In \cref{eq:multnorm}, $\Sigma_{y,\mathbf{x_k}}$ denotes the vector of the correlation coefficients between $y$ and each element of $\mathbf{x_k}$, and $\Sigma_{\mathbf{x_k},\mathbf{x_k}}$ denotes the conformable correlation matrix of the conditioning variables $\mathbf{x_k}$. The conditional variance in \cref{eq:multnorm} corresponds to the MSE of a linear regression model where the variable $y$ is predicted based on the vector $\mathbf{x_k}$. Furthermore, for z-standardized variables, the expression $\Sigma_{y,\mathbf{x_k}}(\Sigma_{\mathbf{x_k},\mathbf{x_k}})^{-1}\Sigma_{y,\mathbf{x_k}}$ corresponds to the $R$-squared of the linear regression of $y$ on $\mathbf{x_k}$, which I denote by $R^2_{y|\mathbf{x_k}}$. The conditional variance in \cref{eq:multnorm} directly speaks to the effect of adding variables in a linear regression model when the data-generating process follows the multivariate normal distribution and $N$ is large enough to essentially reduce estimation error to zero. I will return to a more in-depth discussion of this and other assumptions in \cref{subsec:discussion}.

\section{Returns from Adding Variables in Multivariate Normal Distribution}
\label{sec:results_I}

\subsection{Correlation Distributions with a Mean of Zero}\label{subsec:mean_zero}

I start the exposition with the case in which the expected value of the correlation distribution between variables is equal to zero. In this case, it can be shown that:

\begin{equation}\label{eq:corr_zero}
	E(\sigma^2_{y|\mathbf{x_k}}) = 1 - E[R^2_{y|\mathbf{x_k}}] = 1 - k \, Var(\rho) \, E \Big( \frac{1}{1 - R^2_{y|\mathbf{x_{k-1}}}} \Big),
\end{equation}

\noindent where $k$ denotes the number of variables in $\mathbf{x_k}$ and $Var(\rho)$ is the variance of the correlations. \cref{eq:corr_zero} reveals that the $R$-squared for a conditioning set with $k$ variables depends on the $R$-squared of the conditioning set with $k-1$ variables. Given this recursive structure, it is easy to show that \cref{eq:corr_zero} implies increasing returns to scale in prediction accuracy. \cref{prop:theory} summarizes the results:

\begin{proposition} \label{prop:theory}
	If all variables follow a multivariate normal distribution and the correlation coefficients between variables are independently drawn from a distribution with a mean of zero and finite second moments, the expected conditional variance of the target variable is decreasing and concave in the number of conditioning variables.
\end{proposition}

\cref{prop:theory} implies increasing returns to scale in prediction accuracy under the stated assumptions. The derivation and analysis of \cref{prop:theory}, which is shown in \cref{app:proof}, is greatly simplified by the assumption that the expected correlation between variables is zero. As a result, the average marginal contribution of a new variable in predicting $y$ is not affected by its correlation to other predictors. 

Note that $Var(\rho)$ corresponds to the expected $R$-squared with a single conditioning variable. If the expected $R$-squared were to scale linearly with the number of variables, it should be equal to $k \, Var(\rho)$. Instead, \cref{eq:corr_zero} reveals that the expected $R$-squared obtained with $k$ variables depends positively on the expected $R$-squared obtained with $k-1$ variables. This, together with the fact that the $R$-squared is strictly increasing in $k$, implies that $E[R^2_{y|\mathbf{x_k}}]$ is an increasing convex function, and therefore that $1 - E[R^2_{y|\mathbf{x_k}}]$ is a decreasing concave function.\footnote{Note that $0 \leq E[R^2_{y|\mathbf{x_k}}] \leq 1$ is guaranteed to hold.}

\subsection{Correlation Distributions with a Non-Zero Mean}\label{subsec:mean_nzero}

For the case of correlation distributions with a non-zero mean, I resort to simulations using \cref{alg:simulation}. For a variable vector of dimension $K$ and a distribution $F(\cdot)$ of the correlation coefficients, I simulate $N=100$ learning trajectories. For each learning trajectory, I draw the off-diagonal elements of the correlation matrix from the assumed correlation distribution $F(\cdot)$ and randomly select one target variable $y$. I incrementally increase the number of predictive variables from $k=1$ to $k=K-1$ by adding one new variable in every step of the learning trajectory. Note that I use the method of \cite{higham2002computing} to compute the nearest positive definite correlation matrix if the original realization is not itself positive definite.\footnote{There is a concern that this correction impacts the independence assumption, as the method effectively shrinks large correlations. In terms of the nature of observed returns, I observe no impact of applying the correction, compared to the case where the correction is not needed.} 

\begin{algorithm}
	\caption{Simulation Algorithm for Multivariate Normal Distribution}\label{alg:simulation}
	\begin{algorithmic}[1]
		\State \textbf{set} $K$ and $F(\cdot)$
		\For{$i$ in $1$ to $N$}
		\State \textbf{draw} $K(K-1)/2$ correlation coefficients independently from $F(\cdot)$ and apply
		\Statex \cite{higham2002computing} algorithm to ensure positive definiteness
		\State \textbf{select} $y$ randomly from $\mathbf{x}$
		\State \textbf{set} $\mathbf{x_k} = \emptyset$
		\For{$k$ in $1$ to $K-1$}
		\State \textbf{select} $x_k$ randomly from $\mathbf{x} \backslash (y \cup \mathbf{x_k})$
		\State \textbf{update} $\mathbf{x_k} \gets \mathbf{x_k} \cup x_k$
		\State \textbf{compute} $\sigma_{y|\mathbf{x_k}}^2$ using \cref{eq:multnorm}
		\EndFor
		\EndFor
	\end{algorithmic}
\end{algorithm}

\begin{figure}
	\centering
	\begin{minipage}{.99\linewidth}
		\centering
		\subcaptionbox{\scriptsize $\sigma = 0.100$ \label{subfig:t_01_main_0}}
		{\includegraphics[width=0.22\linewidth]{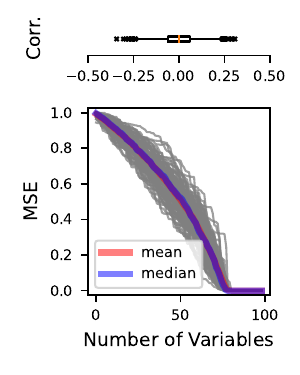}}
		\subcaptionbox{\scriptsize $\sigma = 0.050$ \label{subfig:t_01_main_0}}
		{\includegraphics[width=0.22\linewidth]{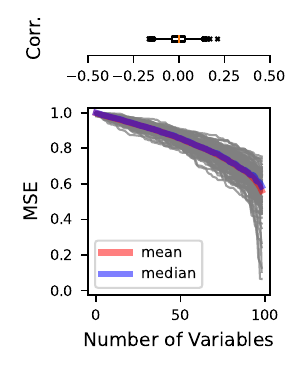}}
		\subcaptionbox{\scriptsize $\sigma = 0.025$ \label{subfig:t_01_main_0}}
		{\includegraphics[width=0.22\linewidth]{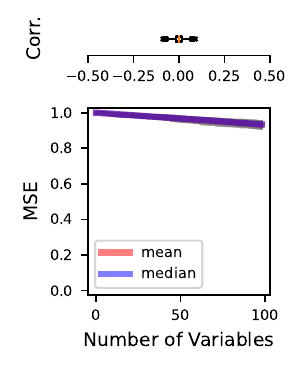}}
		\subcaptionbox{\scriptsize $\sigma = 0.000$ \label{subfig:t_01_main_0}}
		{\includegraphics[width=0.22\linewidth]{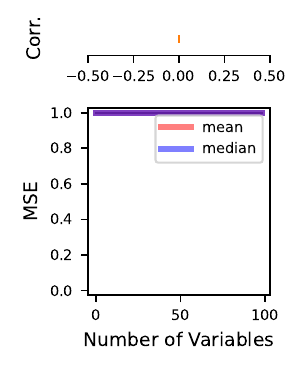}}			
		\caption{\footnotesize Student's t-distributions with a mean of 0 and DoF parameter of 1000 \label{fig:t_main_0}}
	\end{minipage}
	\begin{minipage}{.99\linewidth}
		\centering
		\subcaptionbox{\scriptsize $\sigma = 0.100$ \label{subfig:t_01_main}}
		{\includegraphics[width=0.22\linewidth]{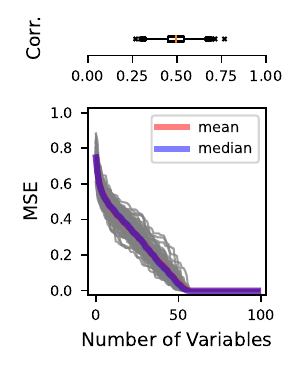}}
		\subcaptionbox{\scriptsize $\sigma = 0.050$ \label{subfig:t_01_main}}
		{\includegraphics[width=0.22\linewidth]{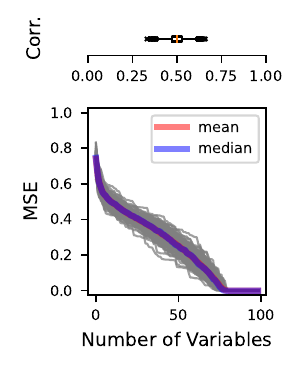}}
		\subcaptionbox{\scriptsize $\sigma = 0.025$ \label{subfig:t_01_main}}
		{\includegraphics[width=0.22\linewidth]{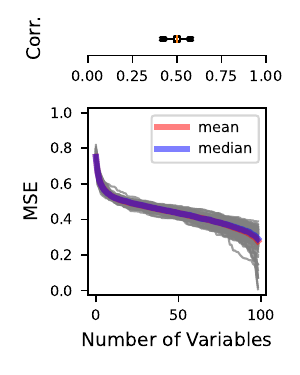}}
		\subcaptionbox{\scriptsize $\sigma = 0.00$ \label{subfig:t_01_main}}
		{\includegraphics[width=0.22\linewidth]{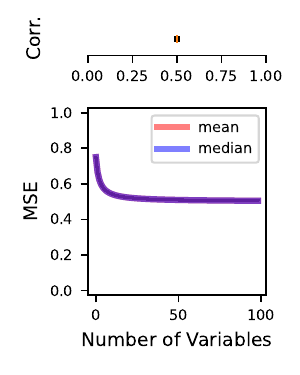}}						
		\caption{\footnotesize Student's t-distributions with a mean of 0.5 and DoF parameter of 1000 \label{fig:t_main}}
	\end{minipage}
\end{figure}

\cref{fig:t_main} displays the results obtained for $K=100$ when sampling the correlation coefficients from a Student's t-distribution with different variances—the mean and the degrees of freedom parameter are set to $0.5$ and $1,000$ (approximating a normal distribution). Each gray line shows one learning trajectory. The boxplots visualize the correlation distribution after applying the method of \cite{higham2002computing}. The results illustrate the main findings obtained for correlation distributions with a mean different from zero: as long as the minimum MSE of zero has not been reached, decreasing returns are generally followed by non-decreasing or increasing returns—except when the correlation distribution is degenerate.\footnote{If the MSE is zero, one or more of the eigenvalues of the correlation matrix, which remains positive definite, are very close to zero. \cref{alg:simulation} accommodates this case by computing the generalized inverse of the correlation matrix.} 

\begin{figure}
	\centering
	\begin{minipage}{.99\linewidth}
		\centering
		\subcaptionbox{\scriptsize $\sigma = 0.100$ \label{subfig:t_01_rob_0}}
		{\includegraphics[width=0.22\linewidth]{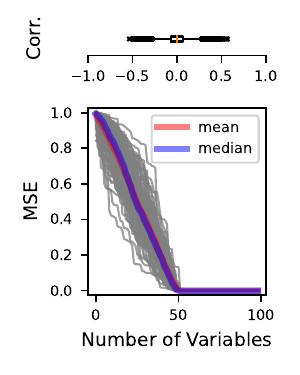}}
		\subcaptionbox{\scriptsize $\sigma = 0.050$ \label{subfig:t_005_rob_0}}
		{\includegraphics[width=0.22\linewidth]{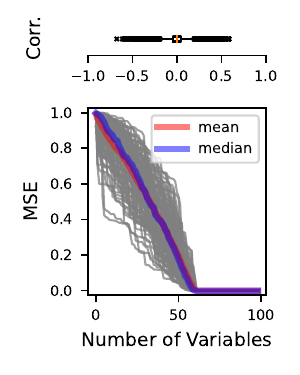}}	
		\subcaptionbox{\scriptsize $\sigma = 0.025$ \label{subfig:t_0.25_rob_0}}
		{\includegraphics[width=0.22\linewidth]{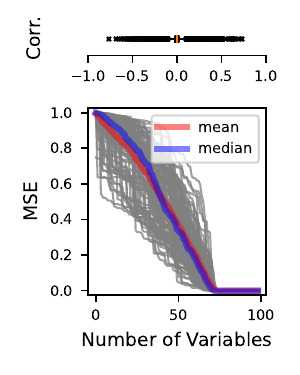}}	
		\subcaptionbox{\scriptsize $\sigma = 0.000$ \label{subfig:t_0_rob_0}}
		{\includegraphics[width=0.22\linewidth]{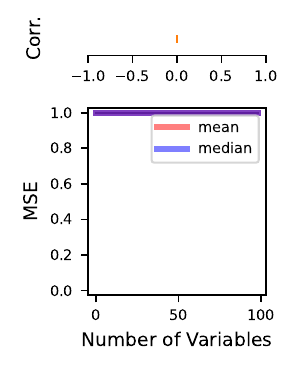}}	
		\caption{\footnotesize Student's t-distributions with a mean of 0 and DoF parameter of 1 \label{fig:t_rob_0}}
		\caption*{\footnotesize \textbf{Note:} The standard deviations refer to the limiting normal distribution.}
	\end{minipage}
	\begin{minipage}{.99\linewidth}
		\centering
		\subcaptionbox{\scriptsize $\sigma = 0.100$ \label{subfig:t_01_rob}}
		{\includegraphics[width=0.22\linewidth]{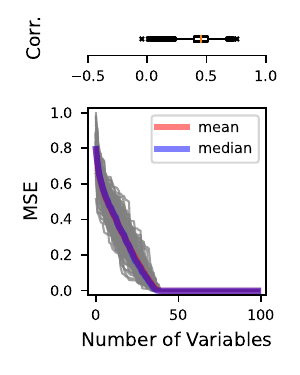}}
		\subcaptionbox{\scriptsize $\sigma = 0.050$ \label{subfig:t_005_rob}}
		{\includegraphics[width=0.22\linewidth]{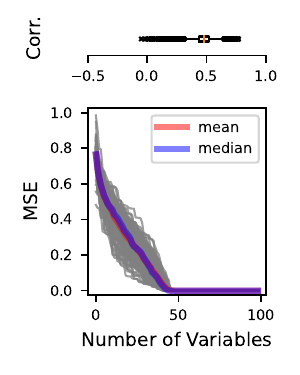}}	
		\subcaptionbox{\scriptsize $\sigma = 0.025$ \label{subfig:t_0.25_rob}}
		{\includegraphics[width=0.22\linewidth]{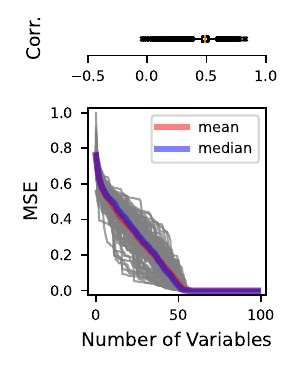}}	
		\subcaptionbox{\scriptsize $\sigma = 0.000$ \label{subfig:t_0_rob}}
		{\includegraphics[width=0.22\linewidth]{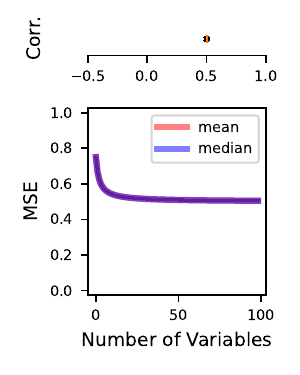}}				
		\caption{\footnotesize Student's t-distributions with a mean of 0.5 and DoF parameter of 1 \label{fig:t_rob}}
		\caption*{\footnotesize \textbf{Note:} The standard deviations refer to the limiting normal distribution.}
	\end{minipage}
\end{figure}

While it proves difficult to obtain exact analytical results for the case of a correlation distribution with a non-zero mean, I now provide a heuristic for the mechanism driving the observed patterns. The argumentation is based on analytically evaluating \cref{eq:multnorm} at the expected value of the correlation (instead of analyzing the expected value of \cref{eq:multnorm} itself) and relies on general properties of multivariate normal distributions.\footnote{The respective formulas and derivations are shown in \cref{app:disc}.} \cref{fig:t_main_0} and \cref{fig:t_main} show the difference in the learning trajectories between correlated and uncorrelated predictors. 

The expected MSE from adding the first variable is $1-E(\rho^2) = 1- Var(\rho) - E(\rho)^2$. When adding a second predictor, the expected collinearity between both predictors will lead to a lesser decrease in the expected MSE.\footnote{\cref{app:disc} provides the exact formula when evaluating \cref{eq:multnorm} at the expectation.} This reflects the phenomenon that, all else equal, adding a predictor correlated with other predictors is less valuable than adding an uncorrelated predictor. As a result of the positive correlation between predictors, we first observe decreasing returns in \cref{fig:t_main}. However, as more variables are added to the model, the collinearity between predictors is projected out. This is reflected by the partial correlations between variables, which approach zero as the number of predictors increases.\footnote{\cref{app:disc} provides the exact formula when evaluating \cref{eq:multnorm} at the expectation.} Thus, for a large number of variables, the collinearity between a new predictor and the already existing set of predictors decreases towards zero. As a result, the negative effect of predictor collinearity vanishes.

In fact, when the expected partial correlation between predictors is zero, the formula to analyze returns should correspond to \cref{eq:corr_zero}, but using the \textit{variance} of the \textit{partial} correlations. This follows from the property that a conditional normal distribution is itself a normal distribution. As a result, we can apply \cref{eq:corr_zero} to the residuals obtained from conditioning out the variables necessary to drive the expected partial correlation to zero (which does not imply its variance is zero). Thus, once the average collinearity between predictors is projected out, the same dynamics as for the case of a correlation distribution with a mean of zero should take over. 

The above heuristic provides an explanation for why we should observe first decreasing and then increasing returns to scale for distributions with a non-zero mean. However, the argument assumes that there remains enough variance in the partial correlations after their average has become zero. If this is not the case, no improvements can be achieved, as, according to \cref{eq:corr_zero}, it is the variance of (partial) correlations that matters. This explains why we observe globally decreasing returns to scale when the correlation distribution is degenerate or nearly degenerate: In this case, after projecting out the collinearity between the predictors, not enough variance remains in order to generate increasing returns to scale.     

\subsection{Discussion}\label{subsec:discussion}

An implicit assumption in the above analysis is that the correlation coefficients between variables are known. In other words, I assume that firms have infinite data and, as a consequence, they can estimate the correlation coefficients without error. I view this as an abstraction for the ability of firms with vast data troves to estimate the parameters of their models with an arbitrary degree of accuracy. One implication of this assumption is that even variables with a very small correlation with the target variable will not be discarded as long as the maximum achievable prediction accuracy has not yet been reached. This is in contrast to a sparse model in which only a few variables have a correlation coefficient large enough to make them worthwhile to be included in the prediction task.  

A sparse model should not significantly alter the insights from the analysis, however. Because variables accumulate in a random order, adding variables with a correlation of zero should have no other effect than rendering the learning curves more ragged, with intervals of no learning followed by sudden learning increases. One way to assess the sensitivity of the results in this direction is to draw correlations from fat-tailed correlation distributions. A fat-tailed correlation distribution corresponds to a scenario where most correlations are close to the mean, with the exception of some large outliers that might suddenly significantly increase the prediction accuracy.     

\cref{fig:t_rob_0} and \cref{fig:t_rob} replicate the main analysis using Student's t-distributions with a degrees of freedom parameter of one. It appears that fat-tailed distributions tend to mitigate, but not eliminate, increasing or non-decreasing returns to scale. Interestingly, individual trajectories might now display decreasing returns to scale, even in the mean zero case, but on average non-decreasing or increasing returns prevail. Another interesting aspect of the results shown in \cref{fig:t_rob_0} and \cref{fig:t_rob} is that fat-tailed distributions might lead to significant differences across trajectories, with some trajectories experiencing most of the gains early on, while other trajectories experience most of the gains very late in the learning process. This shows the possibility that large differences in prediction quality might occur even for the same underlying prediction technology.

Another important assumption is that firms cannot fully control the sequence of variable accumulation. When firms can steer consumers toward variables in a decreasing order of informativeness for a given target, it seems likely that the results presented here could break down under certain conditions. In fact, the individual trajectories with decreasing returns to scale for fat-tailed Student's t-distributions show that this is possible. By contrast, achieving trajectories with decreasing returns to scale seems much harder for thin-tailed distributions. This suggests that the assumption of independently drawn correlations, which effectively implements the random ordering of variables assumption, is not crucial with thin-tails. Although some degree of steering is likely practiced in reality, achieving steering in a perfectly decreasing order of informativeness seems difficult to implement in reality, given idiosyncratic consumer behavior. An additional complication is that steering involves trade-offs because the fastest learning sequence for one target variable will be sub-optimal for another target variable. Since firms seek to achieve good average performance, this further complicates the problem of optimal steering. Consumer steering therefore involves trade-offs which might be an interesting subject for further research.   

Overall, the analysis suggests that increasing returns to scale from additional variables might be common. However, the framework studied so far is very stylized, which raises concerns about the external validity of the results. To address these, I will now empirically analyze returns from additional variables within a more realistic setting. In particular, I will work with data that do not follow a multivariate normal distribution because of the very nature of the variables used. Additionally, my setting will naturally relax the implicit infinite data assumption, and, hence, my approach will be subject to estimation error. Additionally, I will use an algorithm that is more sophisticated than the simple linear regression model that \cref{eq:multnorm} speaks to. 

\section{Returns from Adding Variables in a Collaborative Filtering Algorithm}
\label{sec:results_II}

\subsection{Data}\label{subsec:data}

I use the MovieLens 1M Dataset \citep{harper2015movielens} to analyze returns from additional variables in a more realistic setting. The data contain one million timestamped ratings on a scale from one to five in integer steps. In total, I observe 6,040 users and 3,701 movies. The collaborative filtering algorithm used in this section is trained on past ratings to predict the future ratings of users in a hold-out sample. Note that this setting, in which both the predictors and targets are variables whose roles are interchangeable, corresponds well to the theoretical framework. However, the rating summary statistics in \cref{tab:sum} also reveal a clear departure from the normal distribution assumptions in \cref{sec:model}. The distributions of user activity and item popularity are right-skewed, which is common in the context of recommendation systems: Most users and items are only observed a few times, with outliers who are observed many times more often than the typical user or movie.\footnote{The MovieLens 1M Dataset only includes users with at least 20 ratings.} 

\begin{figure}
	\centering
	\begin{minipage}{0.99\textwidth}
		\centering
		\captionof{table}{Summary Statistics MovieLens 1M Dataset} \label{tab:sum}
		\begin{tabular}{lSccccc}
			\hline
			Variable & {Mean} & Min & p25 & p50 & p75 & Max \\
			\hline
			Rating    & 3.58 & 1   & 3 & 4  & 4 & 5 \\
			Obs. by User   & 165.60  & 20 & 44 & 96 & 208 & 2,314 \\
			Obs. by Movie   & 269.89 & 1 & 33 & 123   & 350  & 3,428  \\
			\hline
		\end{tabular}
		\vspace{0.5cm}
	\end{minipage}
\end{figure}

%

\subsection{The SVD Algorithm}\label{subsec:svd}

The algorithm used in this section is a version of the probabilistic matrix factorization algorithm formally described in \cite{mnih2007probabilistic}. More precisely, I use the so-called SVD algorithm as implemented by the Surprise Python library \citep[see][]{Hug2020}. The difference between the SVD algorithm of the Surprise Python library and the algorithm described in \cite{mnih2007probabilistic} is the presence of user and item biases. More precisely, in the SVD algorithm, the prediction of the rating $r_{ij}$ of user $i$ for movie $j$ is given by:

\begin{equation}\label{eq:svd}
	\hat{r}_{ij} = \alpha + b_i + b_j + f_i' f_j,
\end{equation} 

\noindent where $\alpha$ denotes the constant, $b_i$ denotes the user bias, $b_j$ the movie bias, and $f_i$ and $f_j$ denote the user and item factors, respectively.\footnote{In the probabilistic matrix factorization algorithm formally described in \cite{mnih2007probabilistic}, $\alpha$, $b_i$, and $b_j$ are set to zero.} The parameters are estimated by minimizing the following loss function through stochastic gradient descent:

\begin{equation}\label{eq:svd_loss}
	\mathcal{L} = \sum_i \sum_j \big((r_{ij} - \hat{r}_{ij})^2 + \lambda (b_i^2 + b_j^2 + ||f_i||^2 + ||f_j||^2)\big).
\end{equation}

In \cref{eq:svd_loss}, $\lambda$ denotes the regularization parameter and $||\cdot||$ the L2 norm. Throughout the analysis, I use the default hyperparameters of the algorithm as implemented by the Surprise Python library.\footnote{The details are described in the \href{https://surprise.readthedocs.io/en/stable/matrix_factorization.html\#surprise.prediction_algorithms.matrix_factorization.SVD}{documentation} (last accessed: 27 February 2025).} 

Like in the Singular Value Decomposition, from which the algorithm is inspired, the SVD algorithm seeks to decompose the user-item matrix into user- and item-specific factors whose interactions predict the ratings. The inclusion of bias terms has been found to improve performance in practice \citep{koren2009matrix}. An additional advantage of the inclusion of the bias terms is that it allows for predictions even when the user, item, or both have not been observed in the training data.\footnote{If a user is not observed, the prediction corresponds to the constant plus the item bias. If an item is not observed, the prediction corresponds to the constant plus the user bias. If both are unobserved, the prediction is simply the constant.} However, the key advantage of the SVD algorithm over the conventional Singular Value Decomposition is that it can handle sparse user-item matrices. 

The SVD algorithm constitutes a departure from the theoretical framework, which directly speaks to OLS. However, there are also some commonalities to point out. Importantly, like for OLS, predictions in the SVD algorithm are linear combinations of parameters. Additionally, observing a new movie in the SVD algorithm is associated with increased model complexity and an adjustment of existing parameters. This corresponds well to the theoretical framework, where observing new variables also increases the number of parameters and leads to an adjustment of the already included OLS parameters. In the SVD algorithm, observing a new movie introduces new item factors and leads to an adjustment of existing item and user factors. Therefore, both in the model and the SVD algorithm, predictions are based on linear combinations of parameters, and the acquisition of new variables is associated with higher model complexity and the adjustment of existing parameters.\footnote{While this aspect might appear to apply to all statistical models, it is not always clear how the addition of new variables affects the number of parameters and the adjustment of existing ones. Tree-based algorithms, for example, might not automatically change in their complexity because a variable is added. Rather, the parameters determining model complexity in tree-based algorithms, such as tree depth, the number of trees for Random Forest, or the number of boosting steps for XGBoost, are fixed hyperparameters of the respective algorithms.}

\subsection{Simulation Details}\label{subsec:sim}

I analyze returns to additional variables with SVD-Based Collaborative Filtering using the simulation routine described in \cref{alg:simulation_II}. A direct implementation of the theoretical setting would require holding the number of observations fixed while adding the same variables to all observations. With the MovieLens 1M Dataset, this would amount to adding a set of movies rated by all users to the training data, which is impossible to implement since no single movie has been rated by all users. 

\begin{algorithm}
	\caption{Simulation Algorithm for SVD-Based Collaborative Filtering}\label{alg:simulation_II}
	\begin{algorithmic}[1]
		\small
		\For{$i$ in $1$ to $1,000$}
		\State \textbf{randomly draw} one movie per user for the hold-out sample
		\State \textbf{re-shuffle} the chronological order of observations
		\State \textbf{create} fake timestamp $t$ of newly ordered observations 
		\State \textbf{compute}, for each timestamp, the number $k$ of movies with at least $100$ ratings 
		\For{$k' \in \{100, 200, \ldots, 2,000\}$}
		\State \textbf{update} training data to include all observations with $t \leq \inf\limits_{t} \{t : k \geq k'\}$
		\State \textbf{train} the SVD algorithm on updated training data
		\State \textbf{compute} the RMSE of the SVD predictions on the hold-out sample
		\EndFor
		\EndFor
	\end{algorithmic}
\end{algorithm}

An approximation of the setting described in \cref{sec:model} is to update the model once a certain number of movies has accumulated a sufficient number of observations. This approach acknowledges the large-$N$ assumption by re-evaluating the model performance only when the new movies have accumulated sufficient observations. To implement this idea, I reshuffle the timestamped ordering of movies in each iteration of the outer loop. In the inner loop, I update the training data every time the number of movies with at least $100$ ratings increases by $100$. This procedure emulates real-world data accumulation because, across iterations of the outer loop, more popular movies will reach the $100$ ratings threshold systematically faster. Therefore, \cref{alg:simulation_II} can also be interpreted as simulating system-wide data accumulation scenarios and evaluating the overall system performance at specific checkpoints, where the checkpoints are defined by individual components of the system reaching a critical number of observations. Since there are $2,019$ movies with at least $100$ ratings in the dataset, I stop the inner loop at $2,000$ movies.

To create the hold-out sample, I randomly draw one movie for each user. As a result, the RMSE can be thought of as reflecting the average performance of the algorithm across users, with each user weighted equally. Note that all movies, even those with fewer than $100$ ratings, are used in the training data in each inner step of \cref{alg:simulation_II}. The main motivation for doing so is that excluding movies with fewer ratings would lead to many "unsophisticated" predictions that only include the constant and the user bias in \cref{eq:svd}. By allowing all movies to be included, we minimize the number of such "unsophisticated" predictions and maximize the number of predictions exploiting the full flexibility of \cref{eq:svd}. This mitigates concerns that the learning pattern might be driven by a larger share of sophisticated predictions as the number of movies increases.

\subsection{Results}\label{subsec:simres}
 
\begin{figure}
	\centering
	\begin{minipage}{.99\linewidth}
		\centering
		\subcaptionbox{\scriptsize RMSE \label{subfig:cf_rmse}}
		{\includegraphics[width=0.4\linewidth]{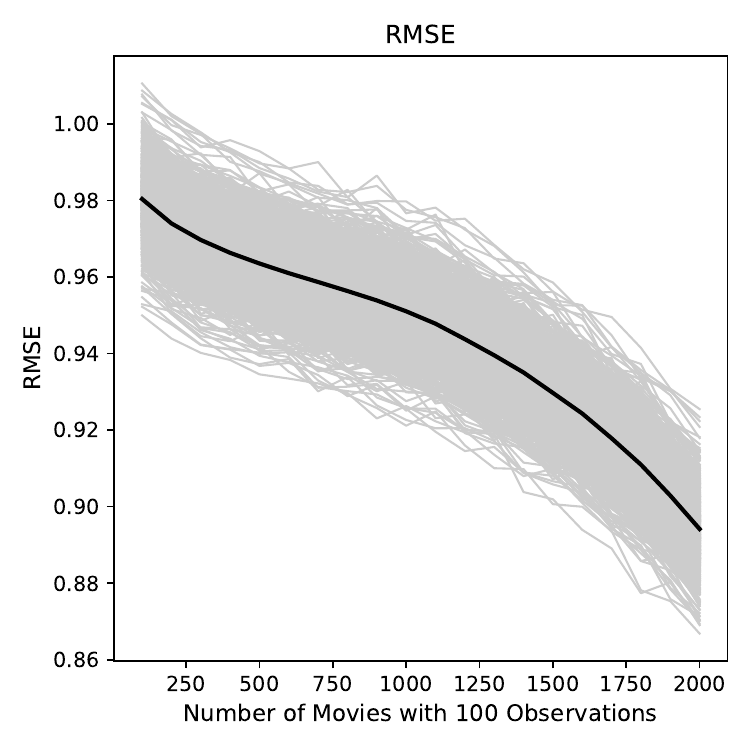}}
		\subcaptionbox{\scriptsize Share Full Model Predictions \label{subfig:cf_sharefm}}
		{\includegraphics[width=0.4\linewidth]{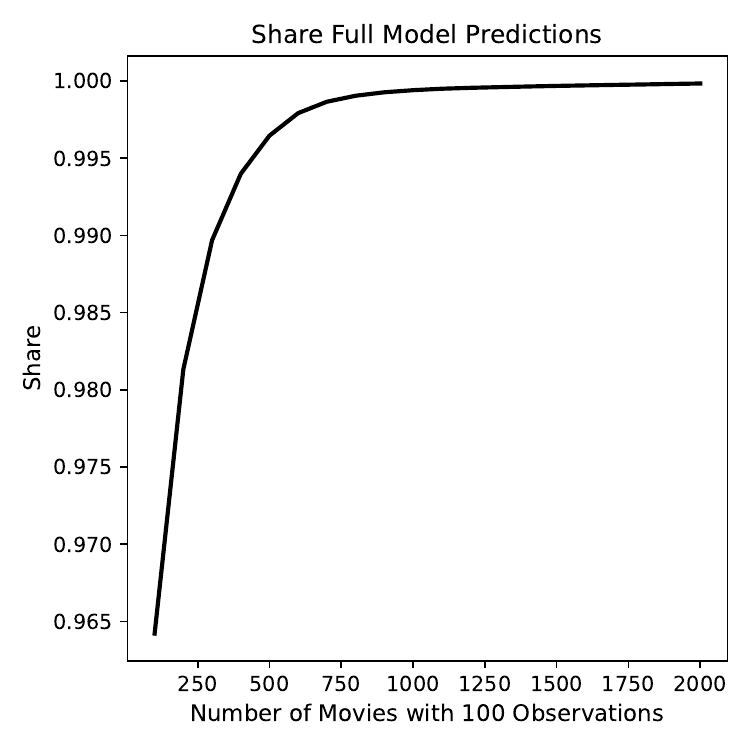}}\vspace{1cm}
			\subcaptionbox{\scriptsize Number of Users and Items \label{subfig:cf_nui}}
	{\includegraphics[width=0.4\linewidth]{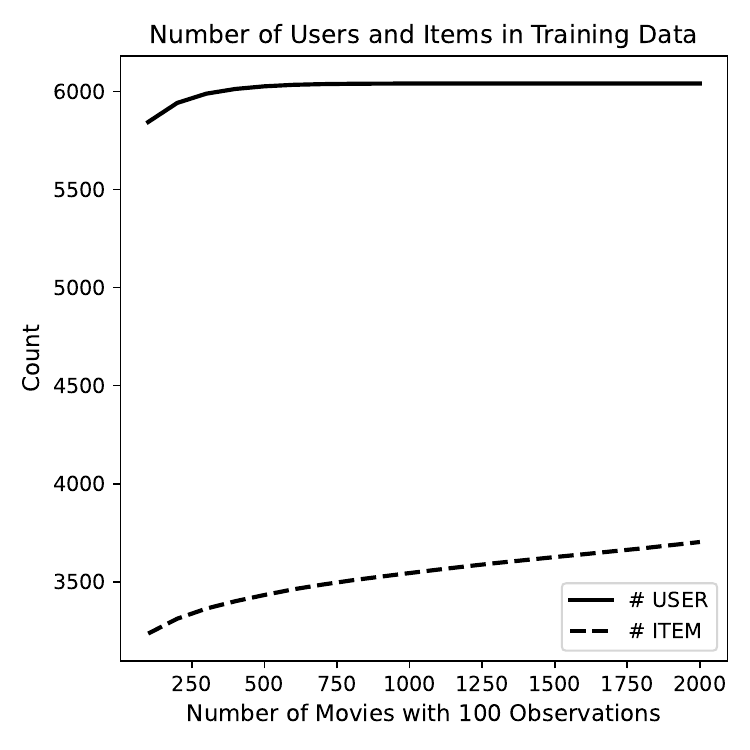}}
	\subcaptionbox{\scriptsize RMSE vs. Root-N Convergence \label{subfig:cf_rmserootn}}
	{\includegraphics[width=0.4\linewidth]{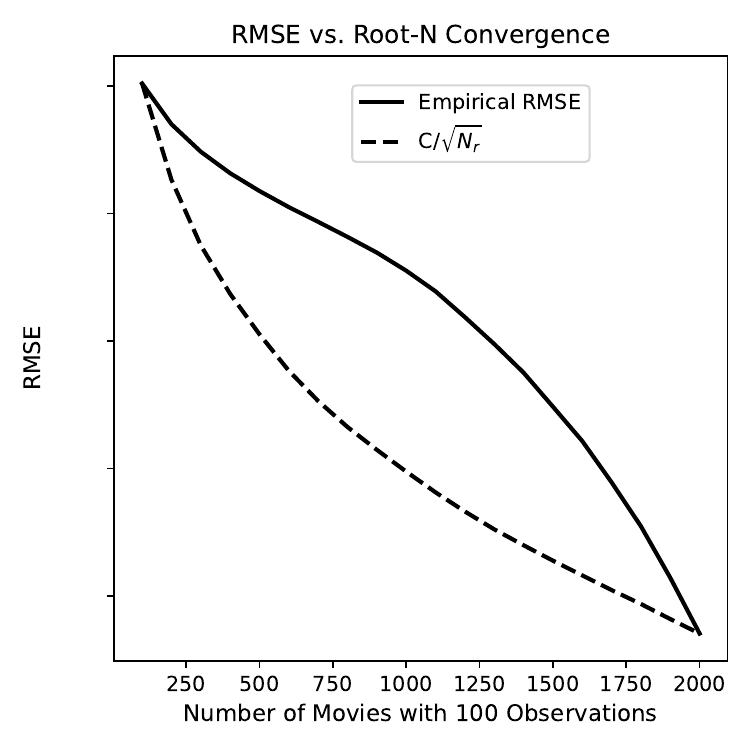}}				
		\caption{\footnotesize Results of SVD-Based Collaborative Filtering Simulations \label{fig:cf}}
		\caption*{\footnotesize \textbf{Note:} In \cref{subfig:cf_rmserootn}, the constant $C$ is set to $100$. The empirical RMSE and the root-N convergence curves are not on the same scale. The number of ratings $N_r$ corresponds to the average across simulations.}
	\end{minipage}
\end{figure}

The results of the simulations are shown in \cref{fig:cf}. Each gray line in \cref{subfig:cf_rmse} shows one RMSE path, while the thick solid line shows the average RMSE path. The results are consistent with the theoretical predictions and show first decreasing and then increasing returns to scale. \cref{subfig:cf_sharefm} shows the average share of observations in the hold-out sample for which the predictions are based on the full model of \cref{eq:svd} (instead of an "unsophisticated" bias-only model). \cref{subfig:cf_nui} shows the average number of users and items in the training sample. From \cref{subfig:cf_sharefm} and \cref{subfig:cf_nui}, it appears unlikely that the pattern of increasing returns to scale in \cref{fig:cf} is driven by compositional changes in the training sample alone. 

Another concern might be that the increasing returns to scale are driven by super-linear increases in the total number of ratings in the training sample as the number of movies increases. In an ideal approximation of the theoretical framework, the total number of ratings in the training sample should scale linearly with the number of movies. This is because, in an ideal approximation, $N_r = K \times N_u$, where $N_r$ is the number of ratings, $K$ is the number of variables (movies), and $N_u$ is the number of users, with the latter being a very large constant number. This is not the case in our analysis, where the number of ratings increases super-linearly with the number of movies. As a result, there is concern that our results could be rationalized by invoking the Central Limit Theorem alone.

To illustrate, recall that the Central Limit Theorem implies that the precision of the estimator increases at a rate proportional to $\sqrt{N}$, where $N$ is the number of samples \citep[see][for a more formal treatment]{bajari2019impact}. If we suppose that $N_r$ is the relevant number of samples to consider, and if $N_r$ scales super-linearly, there is the possibility that $\sqrt{N_r}$ also scales super-linearly. If this is the case, the increasing returns from adding movies might be rationalized by the super-linear increase in $\sqrt{N_r}$ as the number of movies increases. \cref{subfig:cf_rmserootn} contrasts the decrease in prediction error that we empirically observe with the decrease we should expect from the $\sqrt{N_r}$ rule. The dashed line shows how the error of a statistical estimator should decrease with the square root of the number of ratings we empirically observe for a given number of movies. \cref{subfig:cf_rmserootn} shows that despite the super-linear increase in $N_r$, $\sqrt{N_r}$ still increases sub-linearly. As a result, it appears unlikely that our results can be rationalized by invoking the Central Limit Theorem only.

\section{Conclusion}
\label{sec:conc}

This article studies the effect of adding variables on prediction accuracy as measured by the mean-squared-error. Assuming a multivariate normal distribution, it provides a formal proof for increasing returns to scale when the correlations between variables are independently drawn and their mean is zero. When the correlation mean is different from zero, simulations and formal considerations suggest a general pattern of first decreasing and then increasing returns to scale. Simulation results based on more realistic data and a collaborative filtering algorithm provide supportive evidence that the results derived from the theoretical framework hold more broadly.

The findings of this article help rationalize the data hungriness of data-driven firms, which appears inconsistent with statistical theory developed around the $N$-dimension of data that points toward diminishing returns to scale. Instead, increasing returns to scale in the number of variables indicate that large historical databases might provide significant advantages in terms of the prediction accuracy that can be delivered to consumers. As a result, a lack of access to large data troves might constitute a significant barrier to entry when prediction accuracy is important for consumer behavior and their choices.

When training data are easily accessible, increasing returns to scale should raise no competition policy concerns. However, the finding of increasing returns to scale lends credibility to the hypothesis that exclusionary practices with respect to data access can lead to a situation akin to a natural monopoly and should be subject to regulation. Long-discussed policy proposals to address the causes and consequences of the high degree of market power observed in data-driven industries have included mandatory data-sharing and data intermediaries \citep[see][for an overview]{bergemann2023market}. The findings of this article lend support to such proposals, as they support the hypothesis that access to data helps remedy market power—adding to the other benefits inherent in non-exclusionary data access \citep{jones2020nonrivalry}.

While the results from \cref{sec:results_II} show that increasing returns to scale from variables seem to hold more broadly, more work is needed to understand the influence of the selected algorithm. The choice of the SVD algorithm is motivated by its similarity to the model, while simultaneously generalizing it in a meaningful way. Modern-day recommendation systems build on the general framework of the SVD model, but wrap it with more complex functions to map user-item factors into predictions. I leave the empirical investigation of such algorithms for further research.

\newpage
\setlength{\bibsep}{0pt plus 0.3ex}
{\bibliography{subtex/bib}}

\clearpage

\appendix
\section{Appendix}

\subsection{Proof of Proposition I}\label{app:proof}

For exposition, I use numbers to index the target and predictor variables. The index $1$ is used for the target variable, and the indices $2$ to $k+1$ for the $k$ predictor variables. Furthermore, I denote the covariance between variables $i$ and $j$ by $\sigma_{ij}$ and, hence, $\sigma_{ij}^2$ denotes the squared covariance. For consistency, I therefore deviate from standard notation and use $\sigma_{i}$ to denote the variance of variable $i$ (instead of $\sigma_{i}^2$). Furthermore, I denote the expectation of the covariances between variables as $E[\text{cov}]$, and the expectation of the variances as $E[\text{var}]$. Note that the proof does not rely on z-standardized variables.

Using the notation introduced for the proof, the expectation of \cref{eq:multnorm} for non-standardized variables can be written as

\begin{equation}\label{eq:p1}
	E[\sigma_{1|2...k+1}] = E[\text{var}] - E[ \Sigma_{1,2...k+1}(\Sigma_{2...k+1,2...k+1})^{-1}\Sigma_{1,2...k+1}].
\end{equation}

\noindent The expression $E[\text{var}]$ makes clear that the variance of the target variable is itself a random variable. The indices should not be taken as indexing specific variables but rather to designate the target and the order of predictor variables. Note that the inverse of $\Sigma_{2...k+1,2...k+1}$ is the precision matrix. It is easy to verify that the second expectation on the RHS of \cref{eq:p1} can be re-written as:

\begin{equation}\label{eq:p2}
	\sum_{i=2}^{i=k+1} E[\sigma_{1i}^2 \Omega_{ii}] + 2 \sum_{i=2}^{i=k+1}\sum_{j>i} E[\sigma_{1i}\sigma_{1j}\Omega_{ij}],
\end{equation}

\noindent In \cref{eq:p2}, $\Omega_{ii}$ denotes the $i$th diagonal element of the precision matrix and $\Omega_{ij}$ the respective off-diagonal elements. Because the covariances are independent from each other, and because the elements of the precision matrix are only a function of the variables in the conditioning set, we have that $E[\sigma_{1i}^2 \Omega_{ii}] = E[\sigma_{1i}^2] E[\Omega_{ii}]$ and $E[\sigma_{1i}\sigma_{1j}\Omega_{ij}] = E[\sigma_{1i}]E[\sigma_{1j}]E[\Omega_{ij}]$. This can be seen by noting that all elements in $\Omega_{ii}$ and $\Omega_{ij}$ have a leading first index of $2$ or higher, and therefore are independent of all covariances with a leading index of one. Together with the assumption that the covariances have the same distribution, this allows us to re-write \cref{eq:p2} as:

\begin{equation}\label{eq:p3}
	kE[\text{cov}^2]E[\text{diag}(\Omega)] + k(k-1)E[\text{cov}]^2E[\neg \text{diag}(\Omega)].
\end{equation}

\noindent In \cref{eq:p3}, $\text{diag}(\Omega)$ denotes a diagonal element of the precision matrix, and $\neg \text{diag}(\Omega)$ an off-diagonal element. It is easy to see that the second term is zero if the expectation of the covariance distribution is zero. From the properties of the inverse of covariance matrices, $\text{diag}(\Omega)$ corresponds to the conditional variance of one element of the conditioning set given all other elements of the same set. Hence, we obtain:

\begin{equation}\label{eq:p4}
	E[\sigma_{1|2...k+1}] = E[\text{var}] - kE[\text{cov}^2]E\left[\frac{1}{\sigma_{2|3...k+1}}\right]
\end{equation}

Note that \cref{eq:corr_zero} corresponds to \cref{eq:p4} for z-standardized variables. To show that $E[\sigma_{1|2...k+1}]$ decreases at an increasing rate, we now show that $kE[\text{cov}^2]E[1/\sigma_{2|3...k+1}]$ increases at an increasing rate in the number of variables. For $k$ variables, the conditional expectation of the change from adding a new variable, given a realization of the covariance draw, is given by $k\text{cov}^2E[1/\sigma_{2|3...k+1}]$. For $k+1$ variables, this conditional expectation becomes $k\text{cov}^2E[1/\sigma_{1|2...k+1}]$. If $E[1/\sigma_{1|2...k+1}] > E[1/\sigma_{2|3...k+1}]$, the proposition follows because if $\text{cov}^2E[1/\sigma_{1|2...k+1}] > \text{cov}^2E[1/\sigma_{2|3...k+1}]$ holds for every realization of $\text{cov}$, it also holds in expectation. 

Note that $\sigma_{2|3...k+1}$ is the MSE of an OLS regression of $y$ on $k-1$ variables. For every realization of $k-1$ conditioning variables, adding a new variable will strictly decrease the MSE. This is true because the only way that the MSE does not decrease is when the new variable added is collinear with the already included variables or has a covariance of exactly zero with the target variable. Because the covariance matrix is assumed to be positive definite, the first case cannot happen and the second case has a probability mass of zero. The proposition follows.

\subsection{Heuristic for the Case of a Covariance Distribution with a Non-Zero Mean}\label{app:disc}

The goal of this appendix is to argue that \cref{eq:p3} exhibits decreasing returns to scale for small $k$ and increasing returns to scale for larger $k$. The difficulty in proving this property formally stems from the fact that $E[\neg \text{diag}(\Omega)]$ in the second term on the RHS of \cref{eq:p3} is an expression involving the ratio of expectations, which makes it difficult to analyze. More precisely, from general properties of the precision matrix: 

\begin{equation}\label{eq:p5}
	E[\neg \text{diag}(\Omega)] = -E[\rho_{ij|v}\sqrt{(\Omega_{ii}\Omega_{jj})}].
\end{equation}

In \cref{eq:p5}, $\rho_{ij|v}$ denotes the partial correlation between a pair of conditioning variables given all other conditioning variables. The partial correlation is a ratio of lower-order partial correlations, and all terms in expression \cref{eq:p5} are dependent. The intuition for why \cref{eq:p3} exhibits decreasing and then increasing returns to scale in $k$ is that the expected partial correlation diminishes as $k$ increases. As a result, the negative impact of the collinearity between predictors on the explained variance of $y$ (\cref{eq:p3}) decreases as $k$ increases. 

Provided the decay of the partial correlation leads to $E[\neg \text{diag}(\Omega)] \approx 0$ beyond a critical number $k^{\star}$ of conditioning variables, it is easy to make a heuristic argument for why we should have increasing returns above this critical value $k^{\star}$: In this case, we can obtain the residuals of $y$ from a linear regression on $x_{k^{\star}}$ and the residuals of each element of $x_{k'}$ from a linear regression on $x_{k^{\star}}$, where $k'$ denotes conditioning sets larger than $k^{\star}$. Because all residuals follow a normal distribution, \cref{eq:p4} becomes

\begin{equation}\label{eq:p6}
	E[\tilde{\sigma}_{y|x_{k'}}] = 	E[\tilde{\text{var}}] - kE[\tilde{\text{cov}}^2]E\left[\frac{1}{\tilde{\sigma}_{y|x_{k-1}}}\right],
\end{equation}

\noindent where $\tilde{\text{var}}$ is the residual variance of $y$ after a regression on $x_{k^{\star}}$, $\tilde{\text{cov}}$ is the partial correlation between the predictors $x_{k'}$ after their respective regressions on $x_{k^{\star}}$, and $\tilde{\sigma}_{y|x_{k'}}$ is the residual variance of $y$ conditional on the residuals of $x_{k'}$. 

To gain further insights, I evaluate \cref{eq:p3} and \cref{eq:p5} at the averages to understand their behaviors as the effect of randomness vanishes. 

From evaluating \cref{eq:p3} at the averages, we can gain insights into when we should expect decreasing returns to scale when $k$ is small. To simplify the exposition further, I consider the case of a multivariate standard normal distribution, which is without loss of generality. Denote \cref{eq:p3} as a function of $k$ by $\sigma_{\hat{y}}(k)$. The notation reflects that \cref{eq:p3} is the expected variance of the predicted value of $y$ given a number $k$ of conditioning variables. We have $\sigma_{\hat{y}}(k=0) = 0$, $\sigma_{\hat{y}}(k=1) = \overline{\rho}^2$, and $\sigma_{\hat{y}}(k=2) = (2\overline{\rho}^2 - 2\overline{\rho}^3)/(1-\overline{\rho}^2)$, where $\overline{\rho}$ is the expected correlation between the variables of the multivariate standard normal distribution. It is straightforward to verify that the $\sigma_{\hat{y}}(k=2)-\sigma_{\hat{y}}(k=1)$ is smaller than $\sigma_{\hat{y}}(k=1)-\sigma_{\hat{y}}(k=0)$ if $\overline{\rho}<1$. This is the weakest possible upper bound for the expected correlation and indicates that decreasing returns to scale when $k$ is small should be a common phenomenon. 

From evaluating the partial correlation in \cref{eq:p5} at the averages, we can gain an intuition of how fast the partial correlations decrease to zero. Using the recursive property of partial correlations, it is easy to verify that the partial correlation between any pair of conditioning variables when $k\geq2$ is $\overline{\rho}/(1+(k-2)\overline{\rho})$. This quantity vanishes to zero rapidly, which lends credence to the hypothesis that $E[\neg \text{diag}(\Omega)] \approx 0$ for a large enough value of $k$. However, note that a formal full proof would require verifying that the product of partial correlation and the precision matrices in \cref{eq:p5} converges to zero, since the precision matrix is unbounded as $k$ increases; this involves evaluating the expectation of an indeterminate limit.

\end{document}